**Ferroelectric Switching Dynamics of Topological Vortex Domains in a Hexagonal Manganite**

*Myung-Geun Han\*, Yimei Zhu\*, Lijun Wu, Toshihiro Aoki, Vyacheslav Volkov, Xueyun Wang, Seung Chul Chae, Yoon Seok Oh and Sang-Wook Cheong*

[*] Dr. Myung-Geun Han, Dr. Yimei Zhu
Condensed Matter Physics and Materials Science
Brookhaven National Laboratory
Upton, NY 11973, USA
E-mail: mghan@bnl.gov, zhu@bnl.gov
Dr. Lijun Wu
Condensed Matter Physics and Materials Science
Brookhaven National Laboratory
Upton, NY 11973, USA
E-mail: ljwu@bnl.gov
Dr. Toshihiro Aoki
JEOL USA, Inc.
Peabody, MA 01960, USA
E-mail: aoki@jeol.com
Dr. Vyacheslav Volkov
Condensed Matter Physics and Materials Science
Brookhaven National Laboratory
Upton, NY 11973, USA
E-mail: volkov@bnl.gov
Xueyun Wang, Dr. Seung Chul Chae, Dr. Yoon Seok Oh, Prof. Sang-Wook Cheong
Rutgers Center for Emergent Materials and Department of Physics and Astronomy
Rutgers University, Piscataway, NJ 08854, USA
E-mail: sangc@physics.rutgers.edu

Keywords: ferroelectric, topological defects, domain wall, in situ TEM, domain switching

Topological defects that predictably form in a high-temperature disorder phase near its phase transition temperature ($T_c$) persist even afterwards via the Kibble-Zurek mechanism[1-4]. These topological defects are invariant under continuous deformations or perturbations, and thus said to be protected by topology. In condensed matter physics, they often are observable and are believed to play important roles in phase transition[5-7]. Since their recent identification in hexagonal manganites[8-11], these topological defects quickly became a current focus in studies of multifferroics. Even in the presence of severe discontinuities in polarization around the vortex or antivortex core, the formation of these topological defects are observed in hexagonal rare-earth manganites when the crystal is exposed to temperatures above $T_c$[12]. Domain walls in hexagonal manganites also are considered as topologically protected. Two independent research groups reported incomplete poling, resulting in narrow domains[8,10] in hexagonal rare-earth manganite crystals, a feature that was not observed in stereotypical ferroelectrics, such as $PbTiO_3$ and $BaTiO_3$.

Hexagonal $ErMnO_3$ is a geometric ferroelectric since its spontaneous polarization is induced by the structural trimerization of Mn ions and the buckling of Er ions due to mismatch in ionic size between the Re ions and the Mn ions[13-15]. The symmetry-breaking phase transition of the Mn trimerization engenders two possible directions of polarization along the *c*-axis ("+" being parallel to the *c*-axis, and "-" being antiparallel to it ), and three antiphases (α, β, and γ), totaling six distinctive domains (α+, α-, β+, β-, γ+, γ-) as predicted in a single crystalline hexagonal $ErMnO_3$[16]. Recently, Choi *et al.*[8] reported that the walls of the trimerization domain, interlocked with ferroelectric domain walls, emerged from topological defects such, as vortices and antivortices with winding orders, respectively, of α+, β-, γ+, α-, β+, γ- , and α+, γ-, β+, α-, γ+, and β-.

Although such topological defects in hexagonal manganites, such as vortices and domain walls, attracted much attention, their exact roles on the dynamic switching process still are unclear, which could be of great importance for further exploration of novel multiferroics. In this study, using aberration-corrected scanning-transmission electron microscopy (STEM) combining *in-situ* electrical biasing [17-23], we directly observed the unique dynamics of domain switching around a vortex in hexagonal $ErMnO_3$ crystals that we termed "topologically guided partner changing". Six domain walls emerging from a topologically protected and immobile vortex core are paired in a poled state, and each of the three pairs change partners (i.e., neighboring domain walls) in the process of switching to the oppositely poled state. This study establishes a direct relationship between domain wall motions and macroscopic polarization during ferroelectric domain switching of vortex domains in hexagonal rare-earth manganites. Atomic resolution imaging further reveals the atomic topologies across the ferroelectric domain walls interlocked with antiphase boundary.

With ADF STEM-based atomic imaging, we resolved two types of ferroelectric domain wall in ErMnO$_3$, denoted as type-A and type-B walls, respectively shown in **Figures 1a** and **1b**. The [100] projection particularly is useful for imaging the ferroelectric domain walls because spontaneous polarization is easily determined by examining local Er-ion distortions[24-26]. The distortions of the two types of Er-ion columns, Er$_{down}$ and Er$_{up}$, are evident in Figure 1, while the Mn ions along the $c$-axis are almost undistorted. We note that four Er ions shifted upward (parallel to the $c$-axis) while two moved downward (antiparallel to the $c$-axis) in the left region of Figure 1a, yielding an upward net spontaneous polarization (P$_{up}$). The distance (ΔEr) between the Er$_{down}$ and Er$_{up}$ atomic columns from Figure 1a and 1b was measured as 0.510 ± 0.062 Å, viz., slightly larger than the reported value of 0.487 Å based on x-ray measurements[26]. Both domain walls in Figure 1 are the 180°-type because the spontaneous polarizations are antiparallel across them. Some segments of the domain walls are not parallel to the direction of polarization (the $c$-axis), indicating that they either are positively charged (shown in red) in a head-to-head configuration (Figure 1a), or negatively charged (blue) in a tail-to-tail configuration (Figure 1b). Charged domain walls in hexagonal manganites and their unexpected stabilities have been reported by several research groups[9,10,27-29].

An apparent difference between these two types of the domain walls, shown in Figure 1 in the [100] projection, is width of the walls: $\frac{1}{3}$[120] for the type-A wall and $\frac{2}{3}$[120] for type-B wall. Here, we define the width of the domain wall as the separation between two distinct unit-cells from each neighboring domain. In Figures 1c and 1d, the atomic models are depicted in two different projections, i.e., along the $c$- and $a$-axes, for the two kinds of domain walls. For simplicity, we omitted showing the O- and Mn-columns above and below the Er ions in the unit cell. Regardless of the width of the ferroelectric domain walls, we observed that the unit-cell separation near the geometrical centers of all domain walls is always either type A or type B.

**Figure 2** shows schematics of the atomic arrangements near a vortex or antivortex core in the *ab* plane with the integrated structure of APBI + FEBs (solid line), and APBII + FEBs (broken line) alternating around the cores. Here, APB refers to the antiphase boundary, and FEB to the ferroelectric boundary; this model is similar to that proposed by Choi *et al*.[8]. Undoubtedly, the lattice translation symmetries are broken across each domain wall wherein the unit cells are shifted by a vector, $\frac{1}{3}[\bar{1}10]$ in the *ab* plane, corresponding to the relative unit-cell-shift between the two neighboring antiphase (or trimerization) domains. Since the direction of polarization is reversed simultaneously across the domain wall, we can assign a vector of $(\frac{1}{3}[\bar{1}10], \pm)$ for an APBI + FEB or an APBII + FEB (Figure 2a). Here, the minus (plus) sign represents the change in polarization to the direction antiparallel (parallel) to the *c*-axis. Consequently, we determined six domain walls with $(\frac{1}{3}[\bar{1}10], -)$, $(\frac{1}{3}[\bar{1}10], +)$, $(\frac{1}{3}[\bar{1}10], -)$, $(\frac{1}{3}[\bar{1}10], +)$, $(\frac{1}{3}[\bar{1}10], -)$, $(\frac{1}{3}[\bar{1}10], +)$ for a vortex, as indicated with yellow arrows in Figure 2a. The vector sum for the six domain walls emerging from a vortex is $(2[\bar{1}10], 0)$, wherein 0 means there is no change in the direction of polarization. Similarly, for an anitivortex (Figure 2b), the associated six domain walls are assigned as $(-\frac{1}{3}[\bar{1}10], +)$, $(-\frac{1}{3}[\bar{1}10], -)$, $(-\frac{1}{3}[\bar{1}10], +)$, $(-\frac{1}{3}[\bar{1}10], -)$, $(-\frac{1}{3}[\bar{1}10], +)$, $(-\frac{1}{3}[\bar{1}10], -)$, engendering the vector sum of $(-2[\bar{1}10], 0)$. The minus sign in the unit-cell-shift vector $(-\frac{1}{3}[\bar{1}10])$ reflects the reversed winding order of the antivortex compared with that of the vortex. Consequently, it is apparent that a pair, vortex and antivortex, does not result in a net unit-cell-shift as the total vector sums cancel out each other. In Figure 2, we forced the alignment of the domain walls to the [100] direction along which we made our experimental observations. We note that when the walls are viewed along the *a*-axis, two type-A walls and four type-B walls are associated with a vortex, and four type-A walls and two type-B walls with an antivortex. This indicates that a vortex can be distinguished from an antivortex by examining the APB separations around its core along the *a*-axis.

We directly observed switching dynamics near the topological defect (vortex) by applying external electric fields *in-situ* along the *c*-axis (**Figure 3**). We employed the dark-field (DF) TEM imaging method with a large objective aperture including the 020-, 030-, 022-, and 032-spots during switching experiment. We found that this dark-field imaging optimally visualized the domain walls as lines when sample was thick. Additionally, the images showed several thickness fringes that are extraneous to the domain wall observations. In Figure 3, we drew lines for domain walls, which observed in dark-field images (please see Figure S2 in Supplementary Information for the images without the drawn lines). We carried out our series of switching experiments, denoted in alphabetical order, and correspondingly illustrated in Figures 3a to 3m. Domains with parallel polarization to the applied electric field expand, while those with antiparallel polarization shrink, as one can predict for typical ferroelectric domain switching. By measuring the area of $P_{up}$ domains (polarization pointing toward the surface of the sample, or along the *c*-axis), a hysteresis behavior is observed (Figure 3n). For comparison, a polarization (P) – electric field (E) loop electrically measured from a bulk $LuMnO_3$ crystal is shown in Fig. 3(n), which shows larger coercive fields. In fact, it is consistent in that larger field is typically required to achieve a global poling of a bulk $LuMnO_3$ crystal while weaker field is enough to achieve a local poling of a few micron size TEM sample. We note that the three 0 V states (Figures 3a, 3g, and 3m) exhibit a strong preference of $P_{up}$ domains near the surface, which thus suppress the $P_{down}$-dominant remanent state. It indicates the presence of an internal electric field near the surface, locally lowering the energy of the $P_{up}$ domain with respect to that of $P_{down}$ domain. The internal electric field near the surface resulted in significant back switching when negative external field was removed, as can be seen in the domain structure change from Figure 3(l) to 3(m), shifting the P-E loop in Figure 3(n) towards negative voltage side. We attribute this internal electric field to inhomogeneous oxygen vacancy or metallic impurity distributions along the *c*-axis near the surface[30, 31]. Interestingly, we note that the position of paired-walls at the top electrode interface is preserved for all three 0 V states, as depicted by red circles in Figures 3a, 3g, and 3m; this feature is indicative of the restoration of the configuration of the surface domain after removing applied electric fields.

In Figure 3, all TEM images show that the vortex core (marked with a green dot) was fixed during the entire switching process, revealing that its topology protected it. The vortex core where the three up domains and three down domains meet may be electrically neutral and is not influenced by applied electric fields. Also, vortex core can be pinned at defects[32], such as oxygen vacancy, and thus becomes immobile. In addition, the domain walls are closely paired with large electric fields, rather than pair-annihilated as often happens in typical ferroelectric crystals without accompanying antiphase boundaries; examples are PbTiO$_3$ and BaTiO$_3$ wherein a single domain state easily is obtained by electrical poling. The absence of pair-annihilation here can be understood by the partial unit-cell-shift vectors across each domain wall (Figure 2). Around a vortex core, each domain wall carries a unit-cell-shift vector ($\frac{1}{3}[\bar{1}10], -$). For two domain walls paired by an applied electric field, their vector sum becomes ($\frac{2}{3}[\bar{1}10], 0$), i.e., incommensurate with respect to the underlying lattice. The lattice cannot accommodate this partial unit-cell-shift, consequently prohibiting pair-annihilation.

Ferroelectric domain walls tend to align in the direction of polarization so to reduce electrostatic energy owing to discontinuities in the normal component of polarization across domain walls[32]. Significant parts of domain walls in Figure 3 are tilted from the direction of polarization, and thus, are either positively (red) or negatively (blue) charged. **Figures 4a**, **4b**, and **4c** summarize electrostatic interactions between neighboring walls associated with a pair of vortex-antivortex during the switching process, as a half part (vortex) of the vortex-antivortex pair illustrated in Figure 3. We note that a pair of neutral walls and two pairs of oppositely charged walls around a vortex or antivortex are induced by large applied electric fields, as depicted in Figures 3f, 3l, 4b, and 4c. Electric fields maintain the pair of neutral walls parallel to the field, and stabilize the oppositely charged pairs with the aid of the strong electrostatic attraction between neighboring walls. These pairings can be considered as bound states of paired-domain walls since these walls are preserved, especially near the vortex core, even in the absence of applied electric field (Figure 3m). To further investigate the atomic structures of the bound states of paired-domain walls, the sample was thinned after the switching experiment for atomic-resolution STEM. Figure 4d shows a dark-field image of the same vortex after FIB-milling, as studied in the switching experiment (Figure 3) with the final domain structure of Figure 3l (before FIB-milling). The domain structures are slightly relaxed near the new surface but the vortex remained intact during FIB-milling. The tendency toward a $P_{up}$ domain near the surface is mitigated due to our removal of the original surface because the internal electric field caused by point defects likely was present only near the original surface. ADF STEM images taken in the regions of the two type-A walls (orange rectangles) and the two type-B walls (green rectangles) agree with the schematic of the vortex (Figure 2a). The paired neutral walls (Figure 4e) are near perfectly aligned to the *c*-axis; their bound state is relatively stable for a prolonged time of a few months, as shown in Figure 4d. On the other hand, the oppositely charged pairs around a vortex are tilted from the *c*-axis and easily roughened by external perturbations, such as FIB-milling (the two bright domains pointing towards the surface have become widened, Figure 4d). Both bound states carry domains whose width is only about 5 nm (8 unit-cells) for the bound state of neutrally paired walls (Figure 4e), and 0.6 nm (1 unit-cell) for oppositely charged paired walls (top left corner of **Figure 5f**). We found that 8 unit-cells comprise the average width of the walls of the narrow domains inside those bound state of paired walls that do not carry electrostatic charges; this value may be reduced further in oppositely charged paired walls due to their strong electrostatic attraction. We attribute the stability of these narrow domains to the incommensurability of partial unit-cell-shifts across the paired walls that prevents the unification of the bound states, or pair-annihilating even with one unit-cell separation; it probably assures the strong short-range repulsive interaction for the bound states. The one-unit-cell wide domain appears to be close to the vortex core, the atomic structures of which are yet to be resolved.

Based on the observed topologically protected vortex and paired walls with strong electrostatic interactions, we describe the domain switching process in hexagonal $ErMnO_3$ near a vortex as topologically guided partner changing. Considering the immobile vortex core, the domain walls change partners during the switching process via electrostatic interactions. These partner-changing processes are prominent in the switching sequence, especially those depicted in Figure 3c to Figure 3d, from Figure 3f to Figure 3g, and from Figure 3h to Figure 3i therein, the walls' motions are indicated with black arrows.

In summary, we have determined and illustrated that topological defects orchestrate the domain switching process in hexagonal $ErMnO_3$ crystals. With the guidance of the immobile vortex core, domain walls change partners during the switching process to form three bound states of paired-domain walls near a vortex core. The neutrally paired walls were atomically flat and aligned along the *c*-axis, surrounding narrow domains of about 8 unit-cells wide; the oppositely charged paired walls displayed domain widths down to one unit-cell due to strong electrostatic attractive interactions; this is the narrowest ferroelectric domain reported to date. These narrow domains are topologically protected due to the incommensurate sum of the partial unit-cell-shift vectors for each pair of walls, preventing pair-annihilation or their unification. .

*Experimental*

Using a flux method, we fabricated hexagonal ErMnO$_3$ single crystals at 1,200 $^o$C. The crystal was cooled down to room temperature slowly, at the rate of 2 $^o$C/hour. We prepared the TEM sample using a focused-ion-beam (FIB) *in-situ* lift-out technique with 8 keV Ga$^+$ ion energy, finally milling it by low-energy Ar-ions. We placed a movable W probe equipped in a TEM holder (Nanofactory Instruments AB) in contact with the top Pt electrode; we applied various external biases to the ErMnO$_3$ crystal with this electrode grounded (Supplementary Information). A JEOL 2100F Lorentz microscope was used for the *in-situ* electrical biasing experiments. We employed the dark-field (DF) TEM imaging method with a large objective aperture including the 020-, 030-, 022-, and 032-spots to observe ferroelectric domain switching (Figure 3). The DF-TEM images in Figure 3 in the main text were processed with a nonlinear-filter algorithm using Gatan Digital Micrograph software (Gatan Inc.). DF-TEM images after the image processing, used in Figure 3 in the main text, are shown in Figure S2 in Supplementary Information. Annular dark-field (ADF) STEM images of domain walls at the atomic scale were obtained with a JEOL ARM 200F microscope equipped with a spherical-aberration corrector. The images in Figures 1a and 1b in the main text are raw images. The images in Figures 4e and 4f in the main text were deconvoluted by means of maximum entropy (HREM Research Inc.).

*Acknowledgements*

TEM sample preparation in part was carried out by K. Kisslinger at the Center for Functional Nanomaterials, Brookhaven National Laboratory. Authors acknowledge the use of the ARM 200F of JEOL test facility. We thank Y. Horibe for fruitful discussions, V. V. Volkov for helpful TEM image processing, and A. Woodhead for careful reading and editing. Research was carried out, in part, at the Center for Functional Nanomaterials, Brookhaven National Laboratory, supported by the U.S. Department of Energy, Office of Basic Energy Sciences. This work is supported by the U.S. Department of Energy's Office of Basic Energy Science, Division of Materials Science and Engineering, under Contract number DE-AC02-98CH10886. The work at Rutgers was supported by National Science Foundation DMR-11004484.Acknowledgements, general annotations, funding.((Supporting Information is available online from Wiley InterScience or from the author)).

[1]     T. W. B. Kibble, *J. Phys. A-Math. Gen.* **1976**, 9, 1387.

[2]     T. W. B. Kibble, *Phys. Rep.* **1980**, 67, 183.

[3]     W. H. Zurek, *Nature* **1985**, 317, 505.

[4]     W. H. Zurek, *Phys. Rep.* **1996**, 276, 177.

[5]     N. Mermin, *Rev. Mod. Phys.* **1979**, 51, 591.

[6]     D. Fisher, M. Fisher, D. Huse, *Phys. Rev. B* **1991**, 43, 130.

[7]     T. C. Lubensky, D. Pettey, N. Currier, H. Stark, *Phys. Rev. E* **1998**, 57, 610.

[8]     T. Choi, Y. Horibe, H. T. Yi, Y. J. Choi, W. Wu, S.-W. Cheong, *Nature Mater.* **2010**, 9, 253.

[9]     S. C. Chae, Y. Horibe, D. Y. Jeong, S. Rodan, N. Lee, S.-W. Cheong, *P. Natl Acad. Sci. USA* **2010**, 107, 21366.

[10]    T. Jungk, A. Hoffmann, M. Fiebig, E. Soergel, *Appl. Phys. Lett.* **2010**, 97, 012904.

[11]    M. Lilienblum, E. Soergel, M. Fiebig, *J. Appl. Phys.* **2011**, 110, 3623777.

[12]    S. C. Chae, N. Lee, Y. Horibe, M. Tanimura, S. Mori, B. Gao, S. Carr, S.-W. Cheong, *Phys. Rev. Lett.* **2012**, 108, 167603.

[13]    B. B. Van Aken, T. T. M. Palstra, A. Filippetti, N. A. Spaldin, *Nature Mater.* **2004**, 3, 164.

[14]    C. J. Fennie, K. M. Rabe, *Phys. Rev. B* **2005**, 72, 100103.

[15]    D. Y. Cho, J.-Y. Kim, B.-G. Park, K.-J. Rho, J.-H. Park, H.-J. Noh, B. J. Kim, S.-J. Oh, H.-M. Park, J.-S. Ahn, H. Ishibashi, S.-W. Cheong, J. H. Lee, P. Murugavel, T. W. Noh, A. Tanaka, T. Jo, *Phys. Rev. Lett.* **2007**, 98, 217601.

[16]    T. Katsufuji, S. Mori, M. Masaki, Y. Morimoto, N. Yamamoto, H. Takagi, *Phys. Rev. B* **2001**, 64, 104419.

[17]   X. Tan, J. K. Shang, *J. Appl. Phys.* **2004**, 95, 635.

[18]   X. Tan, H. He, J.-K. Shang, *J. Mater. Res.* **2005**, 20, 1641.

[19]   H.-J. Chang, S. V. Kalinin, S. Yang, P. Yu, S. Bhattacharya, P. P. Wu, N. Balke, S. Jesse, L. Q. Chen, R. Ramesh, S. J. Pennycook, A. Borisevich, *J. Appl. Phys.* **2011**, 110, 052014.

[20]   C. T. Nelson, P. Gao, J. R. Jokisaari, C. Heikes, C. Adamo, A. Melville, S.-H. Baek, C. M. Folkman, B. Winchester, Y. Gu, Y. Liu, K. Zhang, E. Wang, J. Li, L.-Q. Chen, C.-B. Eom, D. G. Schlom, X. Pan, *Science* **2011**, 334, 968.

[21]   P. Gao, C. T. Nelson, J. R. Jokisaari, S.-H. Baek, C. W. Bark, Y. Zhang, E. Wang, D. G. Schlom, C.-B. Eom, X. Pan, *Nature Commun.* **2011**, 2, 591.

[22]   Y. Sato, T. Hirayama, Y. Ikuhara, *Phys. Rev. Lett.* **2011**, 107, 187601.

[23]   C. R. Winkler, A. R. Damodaran, J. Karthik, L. W. Martin, M. L. Taheri, *Micron*, **2012**, 43, 1121.

[24]   Q. H. Zhang, L. J. Wang, X. K. Wei, R. C. Yu, L. Gu, A. Hirata, M. W. Chen, C. Q, Jin, Y. Yao, Y. G. Wang, X. F. Duan, *Phys.Rev.B*  **2006**, 85, 020102(R).

[25]   S. C. Abrahams, *Acta Crystallogr.* **2001**, B 57, 485.

[26]   B. B. Van Aken, A. Meetsma, T. T. M. Palstra, *Acta Crystallogr.* **2001**, E 57, i38.

[27]   E. B. Lochocki, S. Park, N. Lee, S.-W. Cheong, W. Wu, *Appl. Phys. Lett.* **2011**, 99, 232901.

[28]   W. Wu, Y. Horibe, N. Lee, S.-W. Cheong, J. Guest,  *Phys. Rev. Lett.* **2012**, 108, 077203.

[29]   D. Meier, J. Seidel, A. Cano, K. Delaney, Y. Kumagai, M. Mostovoy, N. A. Spaldin, R. Ramesh, M. Fiebig,  *Nature Mater.* **2012**, 11, 284.

[30]   J. F. Scott, C. A. Araujo, B. M. Melnick, L. D. McMillan, R. Zuleeg, *J. Appl. Phys.* **1991**, 70, 382.

[31]   M. Dawber, K. M. Rabe, J. F. Scott, *Rev. Mod. Phys*. **2005**, 77, 1083.

[32]  G. Catalan, J. Seidel, R. Ramesh, J. F. Scott, *Rev. Mod. Phys.* **2012**, 84, 119.

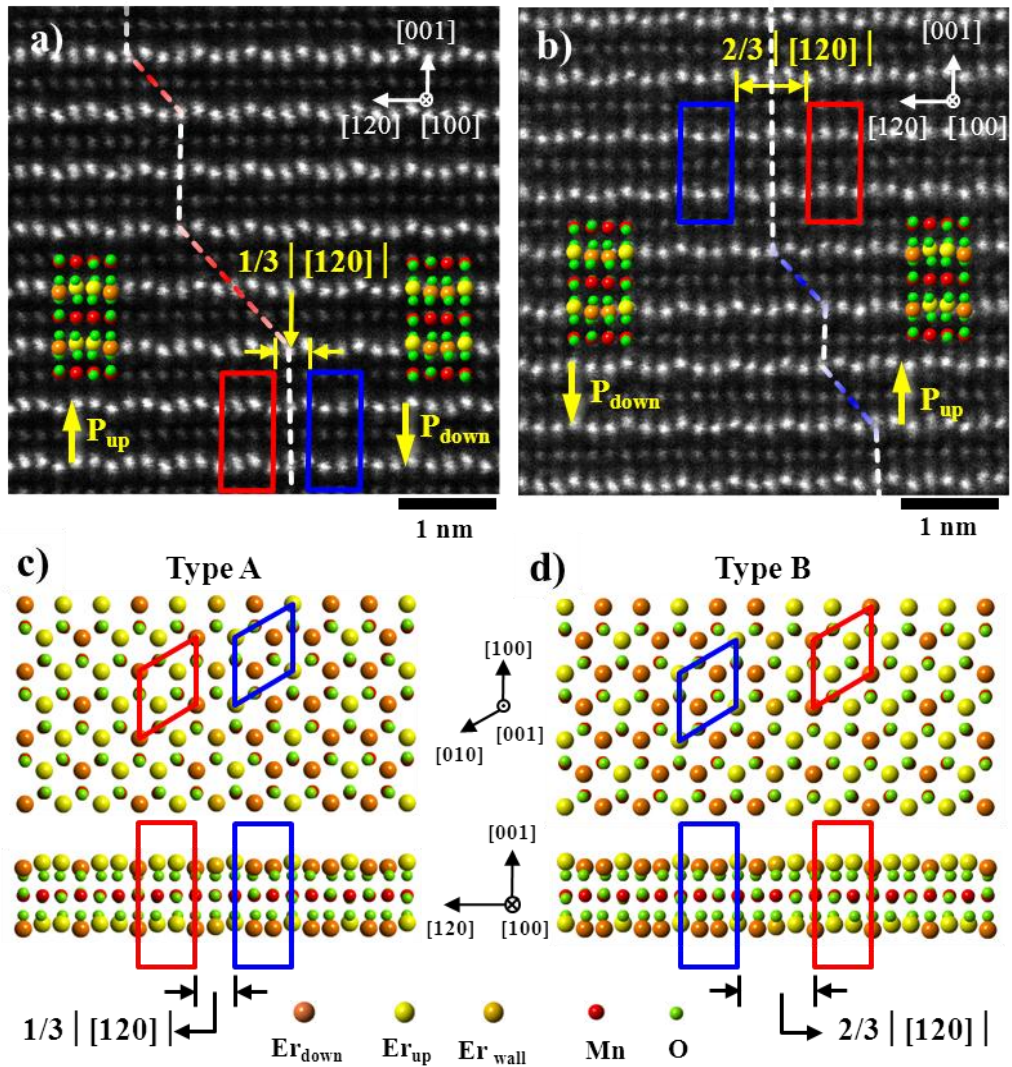

**Figure 1.** Two types of domain walls. ADF-STEM images of type-A (a) and type-B (b) domain walls marked by the broken lines separating two neighboring domains with opposite polarizations. Atomic models and unit cells (rectangles and rhombi indicated in red and blue) also are shown. Charged segments of the domain walls are marked in red (positive charges in a head-to-head configuration) and blue (negative charges in a tail-to-tail configuration). Atomic models of two types of domain walls seen along the [001] (c) and [100] (d) axes. Er ions are either displaced upward (brown, $Er_{up}$), or downward (yellow, $Er_{down}$) along the [001] direction. Er ions located at domain walls are depicted in light brown.

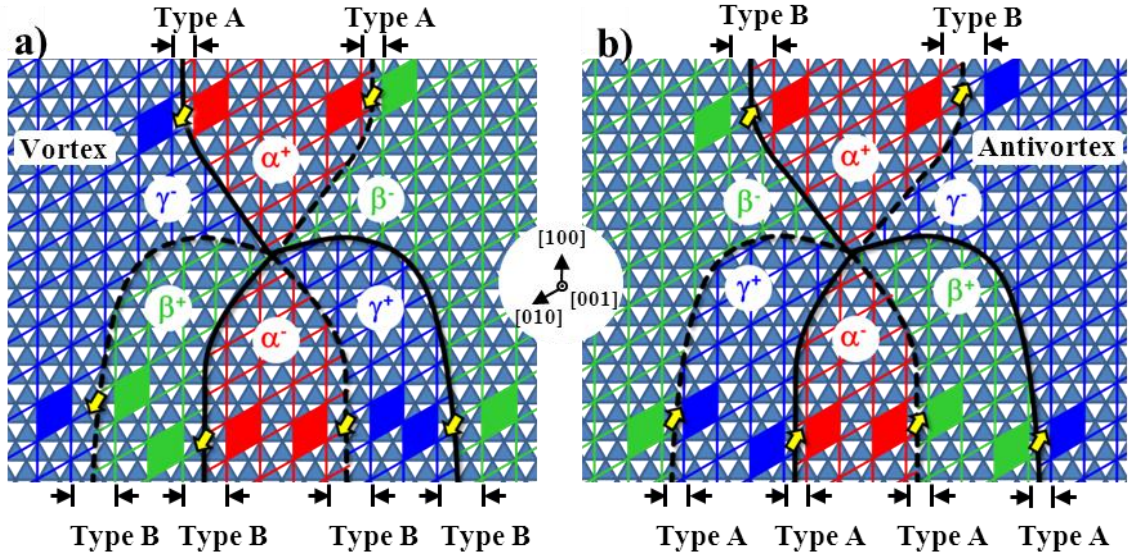

**Figure 2.** Schematics of atomic configurations of domain walls around a vortex a) and antivortex b). Mn ions are located in the center of filled triangles, representing the O-ion bipiramids in the [001] projection. Er ions are located every corner of the filled triangles. APBI (APBII) + FEBs are indicated with solid (broken) lines. Yellow arrows denote the characteristic unit-cell-shift vectors across each wall. The unit-cell-shift vectors for the vortex and antivortex, respectively, are $\frac{1}{3}[\bar{1}10]$ and $\frac{1}{3}[1\bar{1}0]$, and the two type-A walls and the four type-B walls are associated with a vortex, and the four type-A walls and the two type-B walls are associated with an antivortex in these arrangements, as determined by the separations of unit cells across each domain wall.

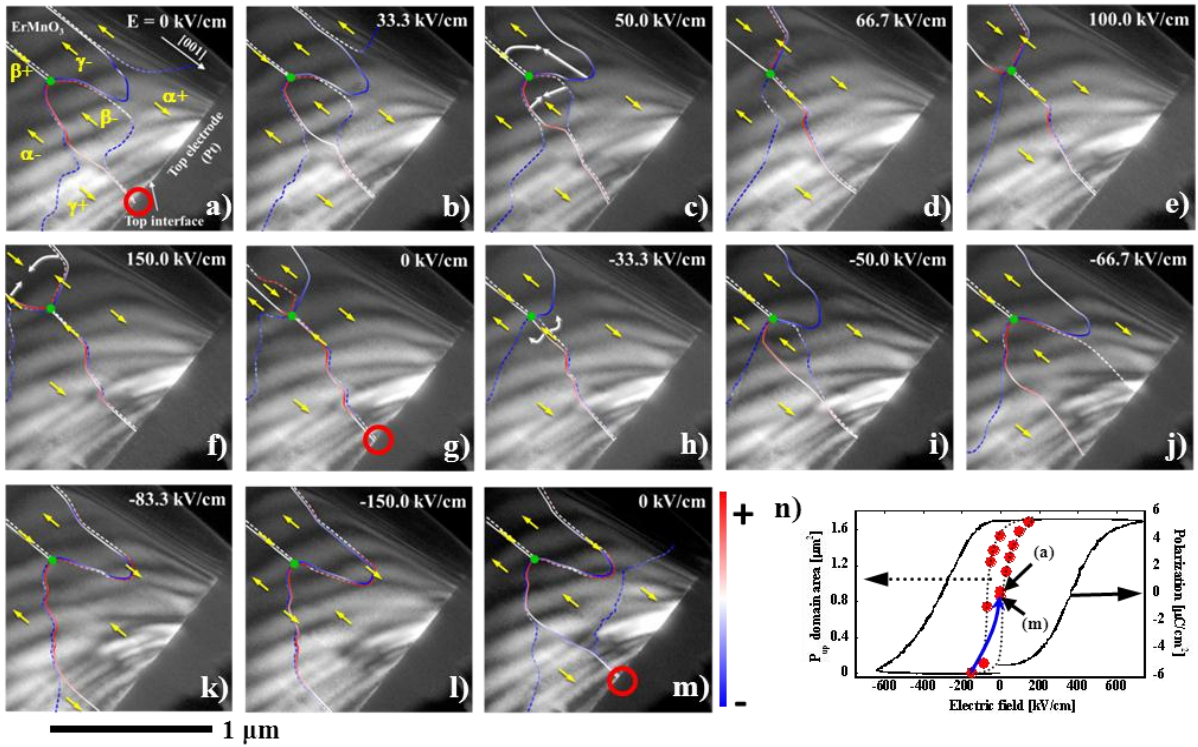

**Figure 3.** Switching dynamics around a vortex. a)-m) Dark-field images showing the order of the switching sequence, denoted alphabetically, with an applied field along the [001] direction. Yellow arrows indicate the polarization direction for each domain. The vortex core is denoted by green dots. Electrostatic charges associated with the domain walls are indicated in red (positive) and blue (negative). The abrupt changes in domain-wall's position from 50 kV/cm to 66.7 kV/cm, from 150 kV/cm to 0 kV/cm, and from -33.3 kV/cm to -50 kV/cm are shown by white arrows. Note that three 0 kV/cm states have similar configurations of the surface domain, indicated by the red circles in a), g), and m). A hysteresis loop (n) was obtained by measuring the $P_{up}$ domain areas for each biased condition represented by red dots. Significant back switching, indicated with the blue arrow (from l to m), is visible. For comparison, a P-E loop electrically measured from a bulk $LuMnO_3$ crystal is also shown.

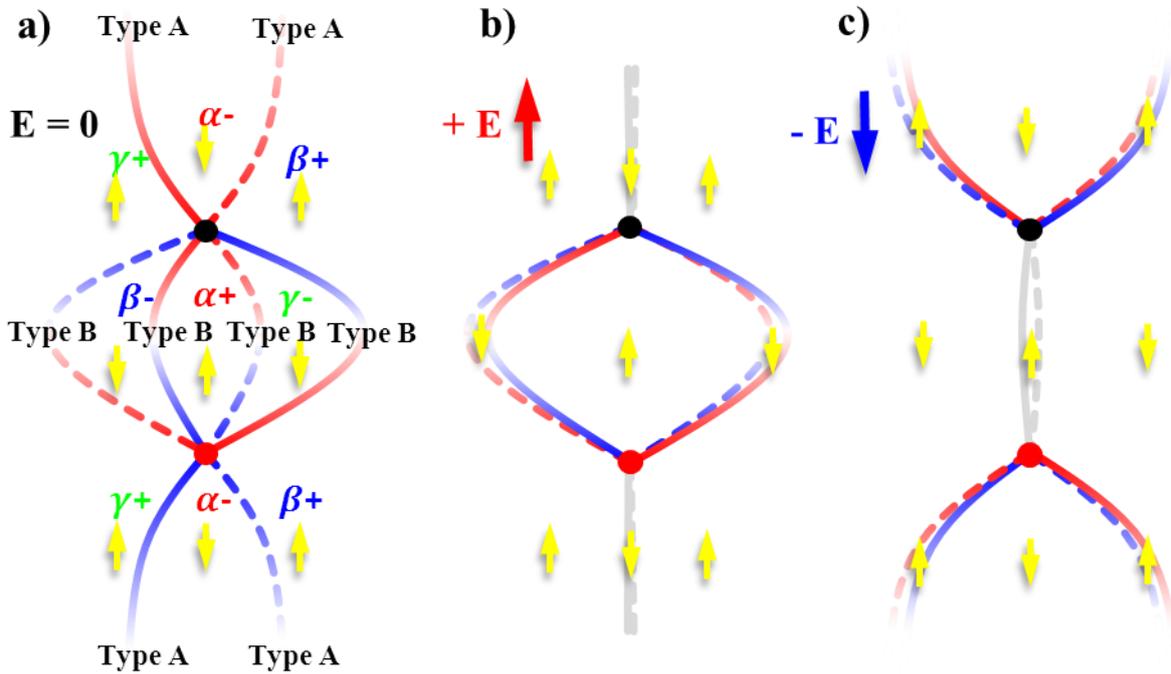

**Figure 4.** Electrostatic interactions between domain walls around a vortex-antivortex pair during switching. a)-c) Unbiased condition (a), saturated state under positive applied field (b), and saturated state under negative applied field (c). The electrostatic charges associated with the domain walls are marked in red (positive) and blue (negative). Similar switching behavior is evident in Figure 3. We note when two domains are paired they carry opposite charges, resulting in a strong attractive interaction.

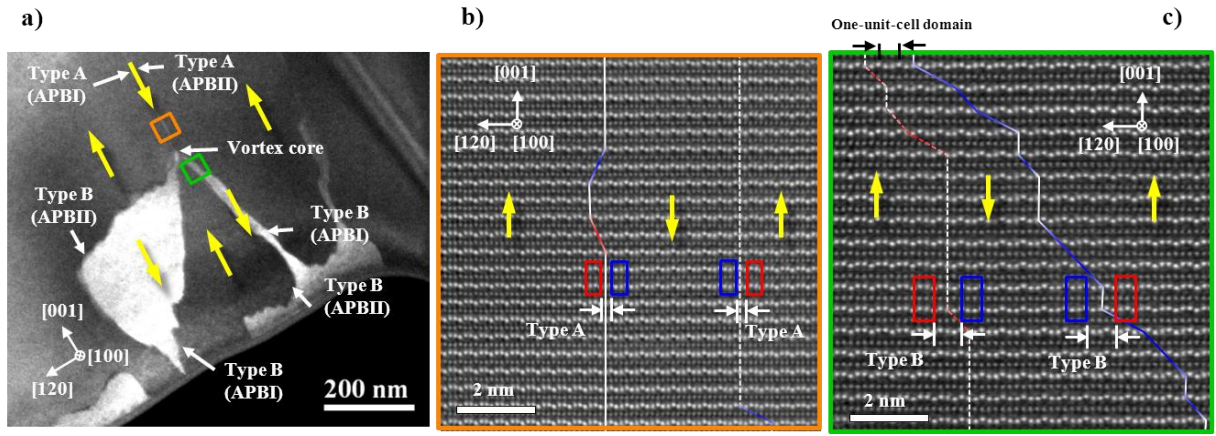

**Figure 5.** Atomic structures of the paired domain walls. a) Dark-field image of a vortex and its associated domain walls obtained with the 001 reflection. b) and c) ADF STEM images from the regions indicated with the orange- and green-rectangles, respectively, in a) are shown in b) and c). Unit cells for each domain near walls are denoted with blue- and red-rectangles. The two walls in b) are type-A walls, and in c) are type-B walls in agreement with the schematic in Figure 2a. Yellow arrows indicate the polarization direction for each domain.

**Supplementary Information**

In **Figure S1a**, a TEM image shows the electrical connection we made for our *in-situ* electrical biasing experiment. We placed a movable W probe equipped in a TEM holder (Nanofactory Instruments AB) in contact with the top Pt electrode; we applied various external biases to the $ErMnO_3$ crystal with this electrode grounded. Figure S1b shows a selected area electron-diffraction pattern from the TEM image of Figure S1a. The applied electric fields clearly were along the [001] direction, as marked with an arrow in Figure 1Sa.

The DF-TEM images in Figure 3 in the main text were processed with a simple sqrt [f] filter algorithm using Gatan Digital Micrograph script (Gatan Inc.). DF-TEM images after the image processing, used in Figure 3 in the main text, are shown in **Figure S2**.

Er displacements were measured by refining the peak positions with respect to the middle position between upward- and downward-displaced Er-columns (i.e., high symmetry position in the paraelectric phase) using computer codes developed by Lijun Wu (Brookhaven National Laboratory), as shown in **Figure S3**. The STEM images were slightly processed to remove noise with a threshold method in frequency space; Fourier transform the images, select only points with a high magnitude (e.g. larger than a threshold level) and finally inverse Fourier transform the images. We determined the peak positions of Er-columns by finding the local maximum intensity from the processed STEM images. For simplicity, we ignored the displacements along the [120] direction, which were much smaller than those along the [001] direction. In **Figure S4**, the line profile of averaged Er-displacements from the frist top five rows (indicated with the yellow box) in Figure S3 B. As seen in the line profile, the Er-displacements is not well defined at the domain wall center (indicated with the arrow). The domain wall width is indicated to seprate two regions with opposite Er-displacement patterns, and thus polarization.

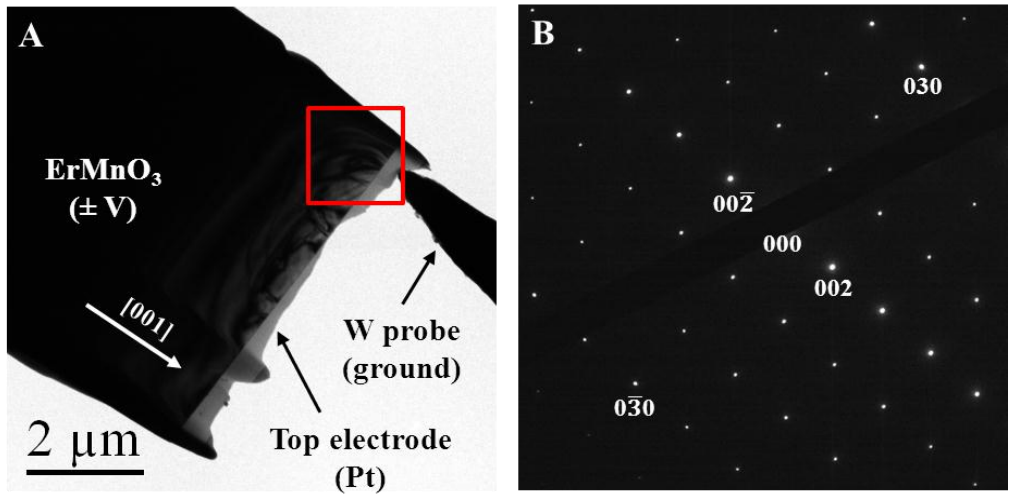

**Figure S1.** a) A TEM image showing the electrical connection for *in-situ* electrical biasing experiments. A movable W probe was placed in contact with the top electrode (Pt) deposited on the *ab* plane of the ErMnO$_3$ crystal. External biases were applied to this crystal while the top electrode was grounded, thereby applying electric fields along the [001] direction. We observed ferroelectric domain switching (Figure 3 in main text) in the area delimited with the red square. b) A selected area electron-diffraction pattern along the [100] projection. Note that the crystallographic [120] direction is equivalent to the 010 reciprocal vector.

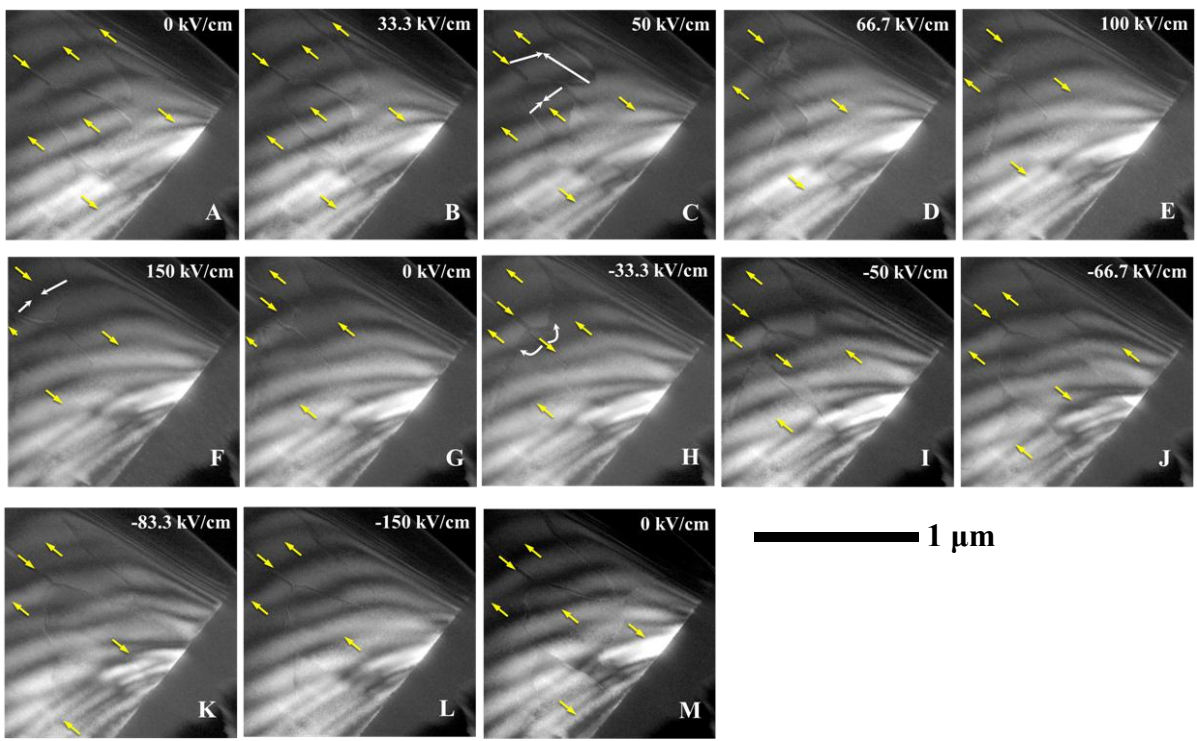

**Figure S2.** a) - m) DF-TEM images after image processing, used in Figure 3 in the main text. Yellow arrows indicate the polarization direction for each domain. The abrupt changes in domain-wall's position are shown by white arrows.

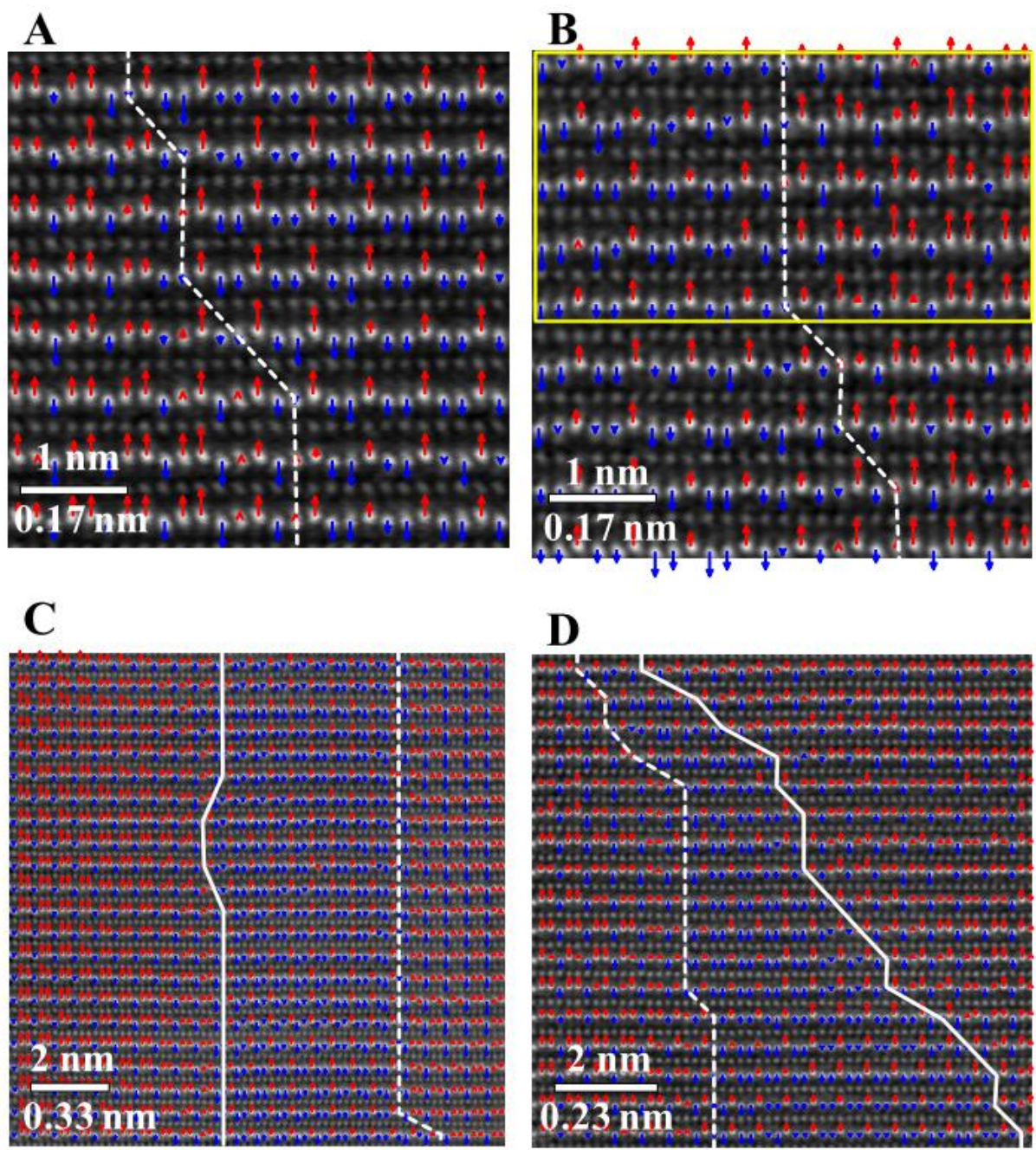

**Figure S3**. a) - d) ADF-STEM images for Figure 1a-b and Figure 5b-c in the main text. The red (blue) arrows for Er-columns indicate the upward (downward) direction. The scale bar shown is both for each image and displacement arrow. The accuracy of displacement was 0.010, 0.013, and 0.020 nm for A and B, C, and D, respectively.

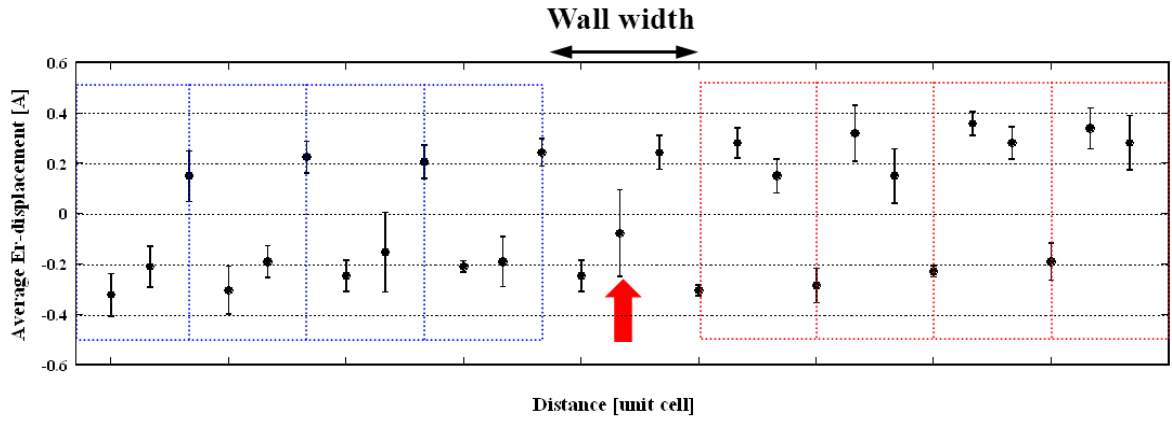

**Figure S4**. Line profile of average Er-displacement across domain walls from the region indicated in the Figure S3 B. Note the Er-displacement at the geometrical center of domain wall marked with a red arrow exhibits a relatively large standard deviation (error bar). Note also that the wall with the finite width separates two regions with the inverted displacement patterns. Blue (red) dotted boxes indicate the unit cells for each region.